# Efficient Microparticle Trapping with Plasmonic Annular Apertures Arrays

Xue Han,[1] Viet Giang Truong,[1*] and Síle Nic Chormaic[1,2]

[1] Light-Matter Interactions Unit, Okinawa Institute of Science and Technology Graduate University, Onna, Okinawa, 904-0495, Japan
[2] Univ. Grenoble Alpes, CNRS, Grenoble INP, Institut Néel, 38000 Grenoble, France

*Corresponding author: v.g.truong@oist.jp

**Abstract**: In this work, we demonstrate trapping of microparticles using a plasmonic tweezers based on arrays of annular apertures. The transmission spectra and the $E$-field distribution are simulated to calibrate the arrays. Theoretically, we observe sharp peaks in the transmission spectra for dipole resonance modes and these are redshifted as the size of the annular aperture is reduced. We also expect an absorption peak at approximately 1,115 µm for the localised plasmon resonance. Using a laser frequency between the two resonances, multiple plasmonic hotspots are created and used to trap and transport micron and submicron particles. Experimentally, we demonstrate trapping of individual 0.5 µm and 1 µm polystyrene particles and particle transportation over the surface of the annular apertures using less than 1.5 mW/µm² incident laser intensity at 980 nm.

## 1. INTRODUCTION

To address the problems associated with conventional optical tweezers for trapping objects in the Rayleigh regime, plasmonic tweezers (PTs) - which can confine the incident laser beam to the nm-scale - have been developed. These devices provide an alternative and robust technique for scaling optical trapping down to subwavelength particles [1-6]. Trapping particles with sizes of 10 nm in diameter or less has been reported [7, 8]. Aside from single nanoparticle trapping, PTs based on large arrays have been used for the optical transport of dielectric micron and sub-micron particles across a chip [9-16]. For example, polystyrene beads with a minimum diameter of 200 nm have been successfully transported using a nano-optical conveyor belt [17, 18]. Another major advantage of PTs is the tunability of the resonance frequency towards the near-infrared (NIR) region. This reduces photodamage and thermal heating of trapped particles, which is particularly important for biological samples. One configuration often considered is the noble metal annular aperture array (AAA) [19, 20]. By reducing the size of the annular aperture, the resonance can be red shifted. Although low incident powers have been shown to efficiently trap particles using PTs, the influence of a strongly enhanced local field at a resonance frequency could lead to rapid damage of trapped particles. The heating effect has been used to assist in optical trapping by increasing micro-fluidic flow, thereby bringing particles into the trapping sites [21, 22]. While heating could be helpful, it should, however, be avoided for most PTs applications.

Recently, self-induced back-action (SIBA) has improved the attainable trap stiffness for lower trapping powers without requiring the use of a laser at a plasmonic resonance frequency [23-27]. A dynamic optical trap, where the long term stability of the trapped particle requires lower average intensity compared to a conventional trap, is achieved by coupling the motion of the particle with the resonance of a nano-aperture.

In our earlier work we demonstrated trapping of nm-sized polystyrene particles (30 nm – 100 nm in diameter) in multiple trapping sites of a plasmonic nanohole array [28, 29]. Here, we present the trapping of micron and submicron particles by using PTs based on arrays of annular apertures. We define one unit of the annular aperture as an inner nanodisk located coaxially inside an outer nanohole (see Fig. 1). These units are connected via nanoslots along the horizontal direction of the array, with nanotips along the vertical direction. Previously, we have shown that such a geometry facilitates a large increase in the transmission coefficient at resonant frequency oscillations [15,16, 28]. Therefore, the plasmonic nanostructure could be used to improve the trapping performance at a low incident trapping intensity when the trapping wavelength approaches the transmission resonance wavelength. In this work, we numerically study the transmission spectra and local $E$-field intensities for AAAs with different sizes of the annular apertures. We observe strong local $E$-field enhancement at the dipole transmission resonance in the annular aperture regions when we use transverse polarisation of the incident light. Experimentally, we demonstrate not only the trapping of dielectric particles using an AAA, but also their transportation across the plasmonic device using a drag force method for a relatively low incident laser intensity of less than 1.5 mW/µm². Our AAA plasmonic nanotweezers have potential as elements in lab-on-a-chip devices for efficient micro- and submicro- particle trapping and transportation with high tunability of the applied laser frequency.

## 2. ANNUALAR APERTURE ARRAY CALIBRATION

We define an array of annular apertures to consist of 10x15 identical apertures on a 50 nm thin gold film. Details of the fabrication procedure are described elsewhere [28]. Scanning electron microscope (SEM) images were used to obtain the dimensions of the fabricated arrays and these values were used in all simulations. Instead of discussing the devices in terms of the size of the annular aperture, for simplicity, we use the inner nanodisk diameter ($d$) of the annular aperture while the outer diameter ($d_{out}$) of the annular aperture is kept fixed. In Fig. 1(a), two SEM images for arrays of $d = 0$ nm and $d = 147.2$ nm are shown. On average, $d_{out} = 293.3 \pm 3.3$ nm, the period $\Lambda = 361.2 \pm 4.6$ nm in both directions, and the width, $w_{slot} = 39.6 \pm 4.6$ nm. The connecting nanoslots are only fabricated along the $x$-direction, leading to the generation of sharp nanotips along the $y$-direction. AAAs with $d = 0$ nm, $128.0 \pm 3.4$ nm, $147.2 \pm 9.7$ nm, $159.2 \pm 4.1$ nm, and $187.3 \pm 1.9$ nm were used for the particle trapping experiments.

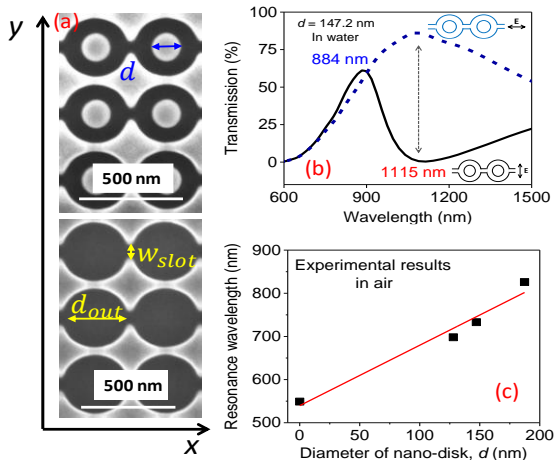

**Figure 1.** (a) SEM images of AAAs with an average $d=147.2$ nm (top) and $d = 0$ nm (bottom). (b) Simulated transmission spectra for the $d=147.2$ nm array using transverse polarisation (black curve) and longitudinal polarisation (blue dash curve). (c) Experimental measurement of the resonance wavelength as a function of the nanodisk diameter. The straight line is a simple linear fit to the experimental data.

Based on the measured dimensions of our fabricated AAAs, we used a finite-difference time-domain (FDTD) method to simulate the transmission spectra and the $E$-field distributions when illuminated by a broadband light source at normal incidence. The simulation unit was 720 nm × 360 nm large, and 1 W of incident power in a sinusoidal pulse over the unit area was considered. The wavelength range was from 300 nm to 2100 nm. The array was modelled as a periodic boundary condition. Figure 1(b) shows the simulated transmission spectra of an AAA with $d = 147.2$ nm for transversally (black solid curve) and longitudinally polarised incident light (blue dash curve) and a cladding of water. For longitudinally polarised light, a broad peak in transmission is observed, whereas for the transverse polarisation, the AAA exhibits a sharp peak at 884 nm in the transmission spectrum. Similar to our previous works, we observed an absorption peak at a wavelength of approximately 1115 nm [28, 29]. This is due to charge accumulation at the sharp edge region of the nanotips. The large transmission difference of ~90% between the two orthogonal polarisations, as indicated by the dashed grey arrow in Fig. 1(b), implies that one can use the designed structure to dramatically improve detection sensitivity using an orthogonal interrogation technique, similar to what has already been reported for an array with different periods along two orthogonal directions so that the symmetry of the nanohole array configuration is broken [30]. A refractive index unit (RIU) sensitivity of approximately $10^{-7}$ was achieved. Our AAAs provide a transmission difference that is almost 10 times higher and it should be possible to reach $10^{-8}$ RIU sensitivity. In order to achieve high field confinement in both the annular apertures and the nanoslots, we select transversely polarised light for the rest of this work.

A microspectrophotometer (CRAIC) was used to measure the transmission spectra of the fabricated arrays. Transmission spectra for AAAs in air, for both simulations and experiments, are shown in Fig. 2 for four arrays of different inner disks sizes. For the simulations, a unit of the annular aperture is illustrated in the top right corner. For experimental results, an image of the array using a regular optical microscope (white light source) is shown in the inset. The transmission resonance peak is identified to show that we have reasonable agreement between the simulations and the experimental results. For larger nanodisks, the resonance is red-shifted and the shift is also apparent in the colour of the inset images. As shown in Fig. 1 (c), there is a linear relationship between the nanodisk diameter and the resonance peak position.

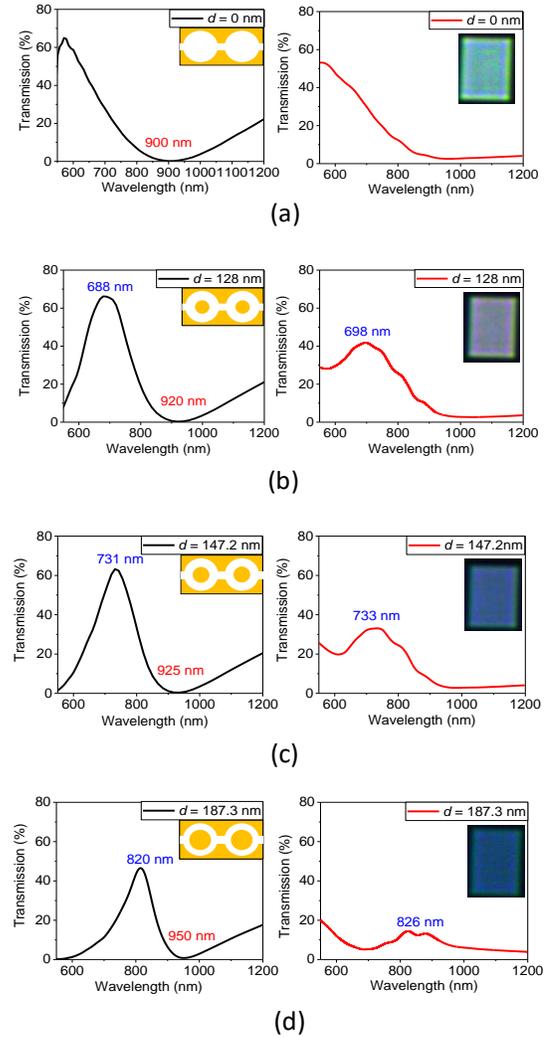

**Figure 2.** Simulated (left) and measured (right) transmission spectra in air for (a) d = 0 nm, (b) d = 128 nm, (c) d = 147.2 nm and (d) d = 187.3 nm. Left insets: The design of each AAA is shown in the inset. Right insets: An image of the fabricated array taken by a regular optical microscope.

The near field distribution was also simulated by examining an interface 10 nm away from the AAA on the cladding layer side. The integrated near field intensity over the whole interface area is studied, as shown in Fig. 3(a). For an array with $d = 0$ nm (black dashed curve), there is no clear evidence of near field enhancement. For the array with $d = 147.2$ nm (red solid curve), a peak is observed at approximately 910 nm, indicating strong near field enhancement around the transmission peak position. Several incident wavelengths are selected for the array with $d = 147.2$ nm in order to demonstrate the distribution of the near field intensity. At 600 nm, the incident light is used to excite the local dipole-like oscillation around the inner nanodisk. At 910 nm, the near field is mainly confined in the aperture area. A strong dipole mode oscillation is generated in the annular aperture cavity. This is due to charges accumulating around the edge regions of the inner disk and the outer hole. This dipole charge resonance oscillation is similar to that of nanoparticles at fundamental plasmonic resonances [31], and we refer to this dipole resonance as the *transmission resonance* from here on. At 1115 nm, the $E$-field is highly localised in the nanoslots and a relatively weak field distribution is observed in the annular aperture region. At 980 nm, hot spots are generated in both the annular apertures and nanoslots, as shown in Fig. 3 (b) & (c). The density of these trapping sites is 18 sites/μm², which is more than 80 times the density of a previous diffraction-limited two dimensional optical lattice created by a holographic technique that was reported in the literature [32], and 4 times higher when compared to similar work on a different type of PT array for micron and nanoparticle trapping and delivery [14]. In practice, this could be beneficial for particle transportation applications. We select this wavelength of 980 nm for the microparticle trapping experiments.

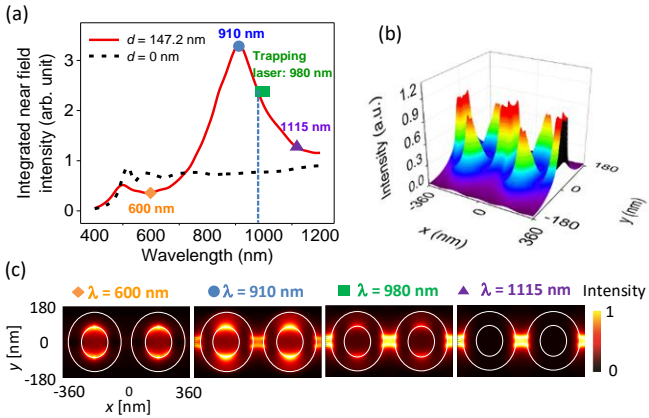

**Figure 3.** (a) Integrated near field intensity versus incident wavelength for an array with $d = 0$ nm (dashed black curve) and $d = 147.2$ nm array (solid red curve). (b) The $E$-field intensity at 980 nm for $d = 147.2$ nm. Multiple hot spots from the annular apertures and nanoslots are observed. (c) The near field intensity distributions for selected wavelengths as labelled in (a) for the array with $d = 147.2$ nm.

## 3. EXPERIMENTAL RESULTS AND DISCUSSION

A Ti: Sapphire laser at 980 nm was selected as the trapping beam. The AAA chip was placed in a sample cuvette with $0.5 \pm 0.1$ μm or $1.0 \pm 0.1$ μm diameter [33] polystyrene (PS) particles in $D_2O$ with a 0.0625% or 0.0125% solid content, respectively. Detergent Tween 20 with 0.1% weight content was used to prevent the formation of clusters. The sample cuvette was mounted on top of a piezo stage. The transmission of the trapping laser through the array was collected by a condenser (50 X, N.A = 0.55) and measured by an avalanche photodetector (APD) at 1 kHz frequency. Direct images of the trapped particles and plasmonic arrays were recorded using a CMOS camera. A pair of annular apertures was fabricated next to the larger array for alignment purposes. First, the incident laser beam was focussed onto this alignment pair of apertures to obtain the focus position along the laser propagation direction, defined as the $z$-axis. By monitoring the transmission in real time, the $z$ position was optimised when a maximum transmission was achieved. Next, the stage holding the sample chamber was moved in the $xy$ plane to focus the incident laser beam onto the AAA. The in-plane position (along the $x$- and $y$-axes) was adjusted to yield maximum transmission, without making any adjustments along $z$. With proper focussing, microparticle trapping was demonstrated for various incident laser powers.

The trap stiffness dependence on the $z$ position is plotted in Fig. 4. For this test, an array with $d = 147.2$ nm was used to trap a 0.5 μm particle. The incident laser beam had a full-width-at-half-maximum (FWHM) of 1 μm and the maximum laser intensity was 1.5 mW/μm². A comparison of the three techniques (power spectral density, equipartition, and drag force methods) used to obtain the trap stiffness is described elsewhere [34].

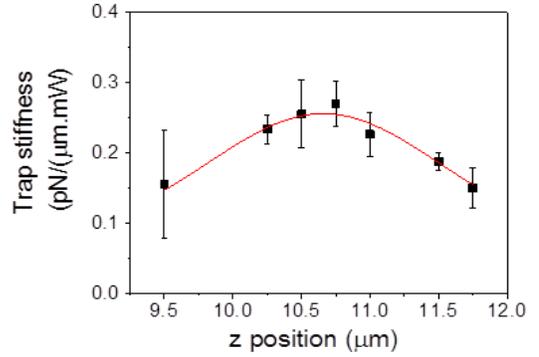

**Figure 4.** Trap stiffness versus $z$ position. The plotted value is the average of two to five measurements and the error bars are the standard deviation. A Gaussian fit is shown by the solid red curve.

### 3.1. Power spectral density (PSD)

For this method, we took the Fourier transform of the transmission of the trapping light versus time in order to obtain the PSD (orange curve). A Lorentzian fit was used to obtain the corner frequency, $f_0$. Results are shown in Fig. 5(a). The trap stiffness, k, is defined by

$$k = 2\pi f_0 \beta, \quad (1)$$

where $\beta$ is the hydrodynamic drag coefficient. We can determine $\beta$ from

$$\beta = 6\pi f_0 \varepsilon(a, h)\mu, \quad (2)$$

where $\varepsilon(a, h) = \frac{9}{15} \ln \frac{h-a}{a} - 0.9588$ is a correction factor, $a$ is the radius of the particle and $h$ is the distance between the centre of the trapped particle and the surface of the device. The value of $(h-a)$ was assumed to be 10 nm for both the 0.5 μm and 1 μm particles.

### 3.2. Equipartition theorem

Recorded images were used to obtain the displacement of the trapped particle from its equilibrium trapped. The centre of the trapped particle was decided for each frame by treating the image of the particle as a circle. The equilibrium position of the trapped particle is the mean value of the central positions of all the frames. The variance, $\langle x^2 \rangle$, is determined by taking the power of the difference of the central position and the calculated equilibrium position, e.g. in the $x$-direction. The trap stiffness is determined from

$$\frac{1}{2} k_B T = \frac{1}{2} k \langle x^2 \rangle \qquad (3)$$

where $K_B$ is Boltzmann's constant and T is the temperature, which we assume to be 300 K. Since the incident power is relatively low at the focal plane of the objective lens and heavy water ($D_2O$) was used to minimise light absorption, we assume that heat generation from the solution around the particle is small enough at really low incident power (less than 1 mW/μm²).

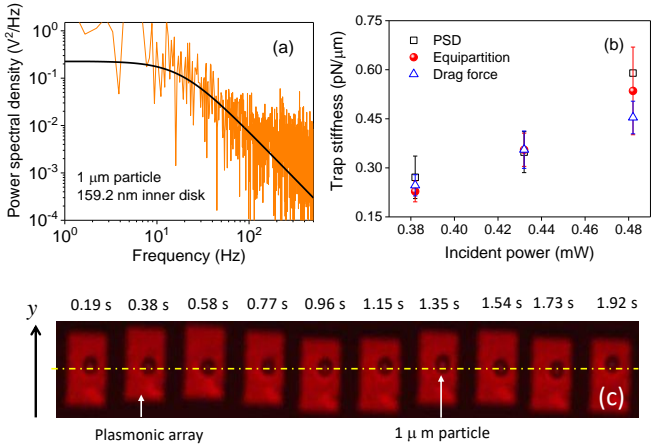

**Figure 5.** (a) Power spectral density curve for a 1 μm PS particle trapped by an array with $d$ = 159.2 nm at 0.6 mW/μm² incident intensity (orange curve). A Lorentzian fit is plotted in black. (b) Plot of trap stiffness versus incident power from the PSD (black squares), equipartition theorem (red spheres), and drag force method (blue triangles). Results are for an array with $d$ = 159.2 nm and a 1 μm particle. Measurements are an average of three or four events for the same incident power and the standard deviation was used as error bars. (c) Images of the AAA with a trapped 1 μm particle on top of it during stage oscillation. The yellow dashed line illustrates the central position of the trapped particle.

### 3.3. Drag force method

Besides examining the trapping performance of the AAA when the piezo stage is stationary, the trap stiffness was also determined when the piezo stage was subjected to sinusoidal motion with a known amplitude, $A_0$, and frequency, $f$. The corner frequency, $f_0$, can be calculated by determining the amplitude of the motion of the trapped particle, $A$ such that

$$f_0 = \sqrt{\frac{A_0 f^2}{A} - f^2} \qquad (4)$$

We assume that, during the stage oscillation, the same strength optical trap is generated at different locations of the plasmonic array. In Fig. 5(c), images of an AAA with a 1 μm particle trapped above it are shown as the piezo stage oscillates along the vertical direction. For trapping experiments, $A_0$ was set to 1 μm and $f$ was set to 1 Hz. The horizontal dashed line shows the position of the trapped particle. It is clear that the array was moving up and down while the oscillation amplitude of the trapped particle was far smaller. This experiment demonstrates that AAA can be used to transport trapped particles over small distances of several micrometres

### 3. 4. Comparison of the three methods

Figure 5 (b) shows the trap stiffness versus the incident power for an array with $d$ = 159.2 nm for a 1 μm particle. We obtained good agreement for the trap stiffness as determined from the three different methods and we see that a higher input power generates a stronger trap. For the input power of 0.5 mW (~ 0.6 mW/μm²), we see a small discrepancy between the values of $k$. This may arise from heating of the gold material. Compared to the drag force method, in the other two methods the trapping laser beam always hits the same spot during the entire trapping period and heat is accumulated – this may affect the trap performance.

### 3.5. Comparison among different AAA

Using the PSD method, we compared the trap stiffness for different AAA. Experimental results for the trap stiffnesses for a 0.5 μm PS particle and a 1 μm PS particle are shown in Fig. 6 (a). For any tested array, we see that the trap stiffness for the 1 μm particle is larger than that for the 0.5 μm particle. We also observe that the strongest trap is obtained when the array consists of disks with $d$ = 147.2 nm for both sizes of particles; the values were 0.25 pN/(μm·mW) for the 0.5 μm and 1.07 pN/(μm·mW) for the 1 μm particles.

The near field intensity for each AAA at 980 nm was studied via simulations and the results are shown in Fig. 6 (a) (right-hand axis). We designate the array with $d$ = 0 nm as $S_1$, $d$ = 147.2 nm as $S_2$, and $d$ = 187.3 nm as $S_3$. We notice that $S_3$, which theoretically provides the highest near field intensity, does not provide the maximum trap stiffness experimentally. Also, the ratio between trap stiffness for the 1 μm ($k_{1μm}$) and the 0.5 μm ($k_{0.5μm}$) particles is not constant for different inner disk diameters, as shown in Fig. 6 (b). For arrays with $d$ = 0 nm, $d$ = 128 nm and $d$ = 187.3 nm, this ratio is approximately 2.6. For an array with $d$ = 147.2 nm ($S_2$), which experimentally provide the strongest trap, the ratio $k_{1μm}/k_{0.5μm} \approx 4.4$ is larger than that obtained for other disk diameters.

To better understand the aforementioned discrepancies between the AAA ($S_2$), which theoretically provides the highest trap stiffness, and AAA ($S_3$), which experimentally provides the strongest near field intensity, we assume that trapping a micron-sized polystyrene particle on top of an AAA changes the refractive index of the cladding layer from $n_{water}$ = 1.33 to an effective refractive index $n_{eff}$, but that this effect does not occur in the nano-apertures. Additionally, to simplify our calculations, we set $n_{eff}$ = 1.57, corresponding to the refractive index of polystyrene. This is a reasonable assumption since the trapped PS microparticle has a size which is comparable to the incident laser beam diameter. With the increased refractive index of the cladding layer, the transmission resonance position is red shifted, as shown in Fig. 6 (c). For $S_1$, this change is relatively small and does not increase the trap stiffness. For $S_3$, this shifts the system out of resonance and is not a desirable effect. Finally, for $S_2$, the resonance peak is shifted closer to the trapping laser beam frequency. This tends to increase the near field intensity, thereby increasing the trap stiffness.

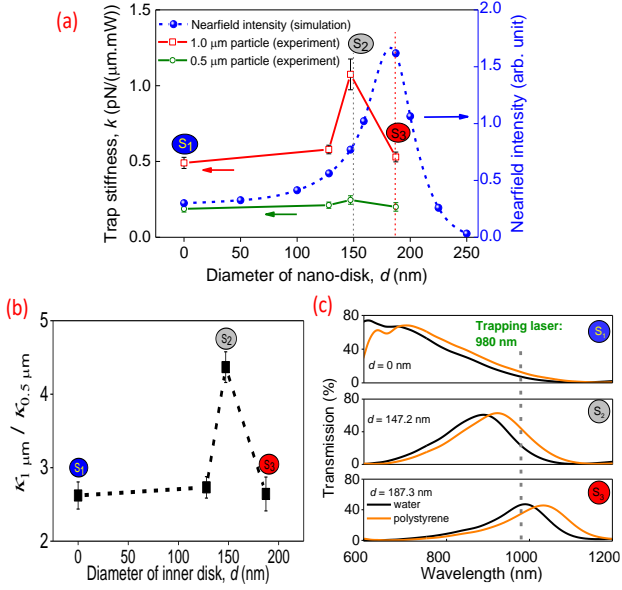

**Figure 6.** (a) Experimental trap stiffness for 0.5 μm (green circles) and 1 μm (red square) polystyrene particles. Error bars are from multiple measurements for different incident powers. Theoretical result of near field intensity (blue line-scatter curve) is shown with the right hand y-axis. (b) Ratio of trap stiffness for 1 μm to 0.5 μm particles for four different (black squares) inner disk diameters. (c) Simulation of the transmission spectra for $S_1$, $S_2$ and $S_3$ with a cladding layer of water and polystyrene.

Based on this observation of the transmission resonance perturbation due to the presence of the trapped particle, we can assume that, with a small redshift of the incident trapping light towards the cavity central wavelength, trapping micron particles on top of the AAA could further improve the trapping performance at relatively low optical intensity.

## 4. DISCUSSION

The above hypothesis of trap stiffness self-adjustment with lower laser intensity excitation for $S_2$ when compared to $S_1$ and $S_3$ is very similar to previous reports on the self-induced back action (SIBA) effect [13,14]. In SIBA trapping, the motion of a trapped particle is dispersively coupled to the resonance optical mode of the system. For a trapping laser slightly redshifted with respect to the resonance wavelength, the nanostructure cavity induces red- and blue shifts toward the laser line when the particle tends to trap and escape from the potential, respectively. Consequently, the particle plays a self-induced dynamic role in such a way that the required average trapping intensity is much weaker when compared to conventional non-resonant trapping regime. It is worth noting that, in a standard plasmonic trapping system, there are certain phenomena that should be addressed for a complete understanding of experimental observations, such as the contribution from heating, thermal convection, and surface roughness effects. Modifications to the surface roughness for different AAA inner disk structures may also cause the particle motion to experience different friction. This could contribute to a change in hydrodynamic interactions between the trapped particle and the device surface which, in turn, could influence the precision of the measured trap stiffness [14].

Here, we emphasise two major issues. Firstly, we had to use a relatively high incident power (up to 1.5 mW/μm² intensity) to trap the 0.5 μm particle, since smaller particles have lower trap stiffnesses for the same incident laser intensity. For the 1 μm particle, an incident intensity of less than 0.6 mW/μm² was used and, therefore, heating should be less of a concern. For the PSD method, additional heating of the gold film for the 0.5 μm particle could increase the Brownian motion of the trapped particle, consequently leading to the higher experimentally observed values of trap stiffness. Therefore, changes in heating on different AAA configurations lead to perturbations on the measurement of the $k_{1μm}/k_{0.5 μm}$ ratio. As shown in Fig. 6 (b), the larger observed ratio for the $S_2$ array could be due to the stronger influence of thermal effects when the trapping laser wavelength approaches the resonance cavity wavelength. Secondly, we note that the observed experimental transmission coefficient is lower when compared to theoretical calculation, as shown in Fig. 2. It is well known that the real and the imaginary parts of the dielectric constant for different AAA structures can differ from Au bulk materials, which we used in our calculation, thereby limiting the accuracy that can be obtained. Furthermore, we note that the $S_3$ array with $d$ = 187.3 nm theoretically provides the highest near field intensity at the interface between the nanostructure and water. However, the larger effective area of the inner disk would also increase the mode volume of the cavity, thereby reducing the SIBA-type effect.

## 5. SUMMARY

Plasmonic tweezers based on AAA were demonstrated. Numerical simulations on the transmission spectra and the near field distributions have been performed. With transversally polarised incident light, we can generate a dipole-like plasmonic transmission resonance in the annular apertures and localised absorption plasmon modes in the nanoslots. To increase the density of trapping sites, we used a laser with a frequency between these two modes. We have experimentally demonstrated trapping of 0.5 μm and 1 μm PS particles and a strong trap stiffness for a low incident laser intensity (less than 1.5 mW/μm²) in the near infrared region. A SIBA-type effect was observed, showing that a back action factor was introduced to boost the trap stiffness when the motion of the trapped particle was coupled to the dipole mode of the annular apertures. A large array of annular apertures, could be integrated into a lab-on-a-chip device to achieve particle trapping and transportation with low incident power and some tunability of the trapping laser frequency.

## 6. FUNDING, ACKNOWLEDGMENTS, AND DISCLOSURES

*Funding*

This work was supported by funding from the Okinawa Institute of Science and Technology Graduate University.

*Acknowledgments*

The authors would like to thank S. P. Mekhail, M. Sergides, I. Gusachenko and M. Ozer for technical assistance and J. Ng and M. Petrov for invaluable discussions.

# 7. REFERENCE

1. M. Righini, P. Ghenuche, S. Cherukulappurath, V. Myroshnychenko, F. J. Garcia de Abajo, and R. Quidant, Nano Lett. 9, 3387-3391 (2009).
2. A. N. Grigorenko, N. W. Roberts, M. R. Dickinson, and Y. Zhang, Nat. Photon 2, 365-370 (2008).
3. Y. Tanaka, and K. Sasaki, Appl. Phys. Lett. 100, 021102 (2012).
4. M. Ploschner, M. Mazilu, T. F. Krauss, and K. Dholakia, J. Nanophoton. 4, 041570 (2010).
5. M. L. Juan, M. Righini, and R. Quidant, " Nat. Photon. 5(6), 348–356 (2011).
6. M. Daly, M. Sergides, and S. Nic Chormaic, Laser Photonics Rev. 9, 309-329 (2015).
7. W. Zhang, L. Huang, and C. Santschi, and O. J. F. Martin, Nano Lett. 10(3), 1006–1011 (2010).
8. A. A. E. Saleh, and J. A. Dionne, Nano Lett. 12(11), 5581–5586 (2012).
9. B. J. Roxworthy, K. D. Ko, A. Kumar, K. H. Fung, E. K. Chow, G. L. Liu, N. X. Fang, and K. C. Toussaint, Jr., Nano Lett. 12, 796-801 (2012).
10. S. H. Wu, N. Huang, E. Jaquay, and M. L. Povinelli, Nano Lett. 16, 5261-5266 (2016).
11. K. Y. Chen, A. T. Lee, C. C. Hung, J. S. Huang, and Y. T. Yang, Nano Lett. 13, 4118-4122 (2013).
12. J. Berthelot, S. S. Acimovic, M. L. Juan, M. P. Kreuzer, J. Renger, and R. Quidant, Nat. Nanotech. 9, 295-299 (2014).
13. V. Garces-Chavez, R. Quidant, P. J. Reece, G. Badenes, L. Torner, and K. Dholakia, Phys. Rev. B 73 (2006).
14. K. Y. Chen, A. T. Chen, C. C Hing, J. S. Huang, and Y. T. Yang, Nano Lett. 13, 4118–4122 (2013).
15. X. Han, V. G. Truong, S. S. Seyed Hejazi, S. Nic Chormaic, Proc. SPIE, 9922, 992227 (2016), revised (2017).
16. X. Han, V. G. Truong, and S. Nic Chormaic, Proc. SPIE 102520R (2017).
17. P. Hansen, Y. Zheng, J. Ryan, and L. Hesselink, Nano Lett. 14(6), 2965–2970 (2014).
18. Y. Zheng, J. Ryan, P. Hansen, Y. T. Cheng, T. J. Lu and L. Hesselink, Nano Lett. 14(6), 2971–2976 (2014).
19. F. I. Baida, A. Belkhir, D. Van Labeke, and O. Lamrous, Phys. Rev. B 74 (2006).
20. Y. J. Liu, G. Y. Si, E. S. Leong, N. Xiang, A. J. Danner, and J. H. Teng, Adv. Mater. 24, OP131-135 (2012).
21. B. J. Roxworthy, and K. C. Toussaint, Jr., Opt. Express 20, 9591-9603 (2012).
22. K. Wang, E. Schonbrun, P. Steinvurzel, and K. B. Crozier, Nat. Commun. 2, 469 (2011).
23. C. Chen, M. L. Juan, Y. Li, G. Maes, G. Borghs, P. Van Dorpe, and R. Quidant, Nano Lett. 12, 125-132 (2012).
24. L. Neumeier, R. Quidant, and D. E. Chang, New J. Phys. 17, 123008 (2015).
25. S. Wheaton, R. M. Gelfand, and R. Gordon, Nat. Photon.. 9, 68-72 (2014).
26. R. A. Jensen, I. C. Huang, O. Chen, J. T. Choy, T. S. Bischof, M. Lončar, and M. G. Bawendi, ACS Photonics 3, 423-427 (2016).
27. P. Padhy, M. A. Zaman, P. Hansen, and L. Hesselink, Opt. Express 25, 26198 (2017).
28. M. Sergides, V. G. Truong, and S. Nic Chormaic, Nanotechnology 27, 365301 (2016).
29. X. Han, V. G. Truong, and S. Nic Chormaic arXiv:1805.01585v1 (2018).
30. A. P. Blanchard-Dionne, L. Guyot, S. Patskovsky, R. Gordon, and M. Meunier, Opt. Express 19, 15041-15046 (2011).
31. U. Kreibig, and M. Vollmer, Optical Properies of Metal Clusters, Springer Series in Materials Science, Berlin (1995).
32. K. Ladavac, K. Kasza, and D. G Grier, Phys. Rev. E, 70, 010901, (2004).
33. https://www.sigmaaldrich.com/Graphics/COfAInfo/SigmaSAPQM/SPEC/L2/L2778/L2778-BULK______SIGMA____.pdf
https://www.sigmaaldrich.com/Graphics/COfAInfo/SigmaSAPQM/SPEC/L3/L3280/L3280-BULK______SIGMA____.pdf
34. K. C. Neuman, and S. M. Block, Rev. Sci. Instrum. 75, 2787-2809 (2004).